\documentclass[twocolumn,showpacs,amsmath,amssymb]{revtex4}
\usepackage{dcolumn}
\usepackage{graphicx}

\begin{document}

\title{Exact solutions for equilibrium configurations
of charged conducting liquid jets}

\author{Nikolay M. Zubarev}
\email{nick@ami.uran.ru}
\author{Olga V. Zubareva}

\affiliation{Institute of Electrophysics, Ural Branch, Russian
Academy of Sciences,\\ 106 Amundsen Street, 620016 Ekaterinburg, Russia}

\begin{abstract}
A wide class of exact solutions is obtained for the problem of finding the
equilibrium configurations of charged jets of a conducting liquid; these
configurations correspond to the finite-amplitude azimuthal deformations
of the surface of a round jet. A critical value of the linear electric
charge density is determined, for which the jet surface becomes
self-intersecting, and the jet splits into two. It exceeds the density
value required for the excitation of the linear azimuthal instability of
the round jet. Hence, there exists a range of linear charge density
values, where our solutions may be stable with respect to small azimuthal
perturbations.

\end{abstract}


\maketitle

\section{Introduction}

Cylindrical jets are known to be unstable with respect to small surface
perturbations because of the development of the Rayleigh instability
caused by capillary effects \cite{ray}. For electrically charged jets,
electrostatic forces are an additional factor that determines the system
behavior. The Coulomb interaction of electric charges suppresses the
large-scale axial capillary instability. On the contrary, it can lead to
the growth of nonaxisymmetric modes of disturbances which are stable for
the uncharged jet (see \cite{mel,sav,gro} and references therein). 
In order to understand the main laws governing the behavior of a charged
jet, it is important to define conditions when the mutual compensation of
the electrostatic and capillary forces is possible, as well as the
conditions when such compensation is impossible. Therefore, the necessity
arises to determine the region of existence of stable solutions for the
problem of the equilibrium configurations of the jet surface.

In this paper we consider possible equilibrium shapes of a charged
infinite jet of a conducting liquid (an electric charge distributes itself
over the liquid surface so that the electric field potential is constant
everywhere inside of the conductor). In doing so, we restrict our analysis
to the particular case of the azimuthal deformations of the initially
circular jet (axial deformations can be suppressed by the longitudinal
magnetic field). For this case, the instability is induced by the
electrostatic forces, while the surface tension forces play a stabilizing
role.  

We are presently only aware of a few nontrivial solutions of the classical
problem in electrostatics, that is the problem of finding the stationary
configurations of the charged surface of a conducting liquid (the flat
surface, the circular cylinder surface, and the sphere belong to the
trivial solutions). We should primarily mention the so-called Taylor cone.
In Ref.~\cite{tay}, Taylor has demonstrated that the surface electrostatic
pressure for a cone with angle $98.6^{\circ}$ is inversely proportional to
the distance from its axis and, hence, can be counterbalanced by the
capillary pressure (the sole exception is the cone apex where the force
balance condition is violated). Recently, the first author of the present
paper has shown \cite{zu1} that for the case of plane symmetry, the
problem of the steady-state shape of the free surface of a conducting
liquid in an external electrical field is mathematically similar to the
problem of the progressive capillary wave solved by Crapper \cite{cra}.
The analogy provides an easy way of solving the electrostatics problem
\cite{zu2}. Finally, the method of constructing exact solutions for the
equilibrium configurations of the charged two-dimensional drops was
proposed in Ref.~\cite{zu3}. It should be noted that apart from the
transformations, this problem coincides with that of the stationary 
shape of a two-dimensional air bubble in a circulatory ambient flow. The
solutions found by Zubarev for an arbitrary mode number $n$ \cite{zu3}
were independently given by Crowdy for $n=2$ \cite{cro} and by Wegmann and
Crowdy for $n=3,4,5...$ \cite{weg} in the problem with the bubbles. The
approaches developed in Refs.~\cite{zu3,cro,weg} turn out to be useful for
the following analysis of possible configurations of charged jets.

The article is made up as follows. In section~II we give the equations
defining the equilibrium configuration of the charged surface of a
conducting liquid for the case of plane symmetry. It is shown that
conformal transformations allows us to reduce the investigation to the
analysis of a nonlinear boundary-value problem on a half-plane for the
Laplace equation. In section~III, using the results of
Refs.~\cite{zu3,cro,weg}, we obtain exact solutions for the jet
configurations corresponding to the azimuthal mode numbers $n=2,3,4...$
In section~IV the equilibrium surfaces corresponding to our solutions are
investigated. We formulate the conditions under which the surfaces become
self-intersecting, and the jet splits into several separate jets. In
section~V we analyze the dependence of the jet surface deformation
amplitude on the control parameter (linear electric charge density) for
different azimuthal numbers. It turns out that we deal with a soft loss of
stability of the round jet surface for $n=2,3,4$ (supercritical
bifurcation) and with a hard loss of stability for $n>4$ (subcritical
bifurcation). Section~VI contains our conclusions and some remarks
concerning conditions whereby the solutions obtained can play an important
role in the jet behavior.

\section{Initial equations}

\begin{figure}
\includegraphics{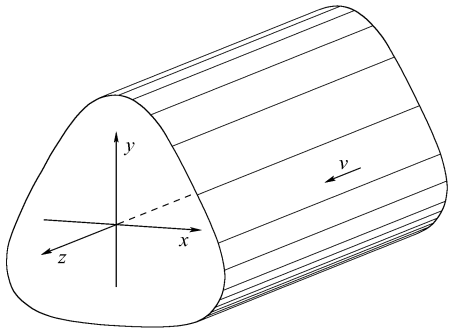}
\caption{\label{fig:fig_1}
The geometry of the azimuthally perturbed liquid jet is represented
schematically. The jet cross-section is constant along the $z$-axis.
}
\end{figure}

Let us write the equations of electrostatics that describe a stationary
profile of the charged surface of the conducting liquid jet with constant
cross-section along the direction of its motion (see
Fig.~\ref{fig:fig_1}). The distribution of the electric field potential
$\varphi$ in the plane of the jet cross-section $\{x,y\}$ is determined by
the Laplace equation: 
$$
\varphi_{xx}+\varphi_{yy}=0.
$$
It should be solved together with the condition that the conductor surface 
is equipotential, $$\varphi=0,$$ and also the condition that the field of
the charged conductor coincides at infinity with the field of a uniformly
charged straight filament: 
\begin{equation}
\varphi\to-Q\ln(x^2+y^2), \qquad x^2+y^2\to\infty,
\label{stac3}\end{equation}
where $Q$ is the linear electric charge density of the conductor.

The jet will be considered to move at the constant velocity $v$ along the
$z$-axis of the Cartesian coordinate system (a fluid is at rest in the
system of coordinates moving with the jet). Then the equilibrium relief
of the charged boundary of a conducting liquid is determined by the 
Laplace-Young stress condition, that is the balance condition for the 
electrostatic and capillary forces acting on the surface:
\begin{equation}
(8\pi)^{-1}(\nabla\varphi)^2_{\varphi=0}+\alpha C+p=0,
\qquad \varphi=0,
\label{stac4}\end{equation}
where $\alpha$ is the surface tension coefficient, and
$C$ is the local curvature of the surface. The constant $p$ is expressed
in terms of the jet velocity ($v$), liquid density ($\rho$), and the
external ($p_e$) and internal ($p_i$) pressures: 
$$
p=p_i-p_e-\rho v^2/2.
$$

It is apparent that an infinitely long cylindrical jet with a circular
cross-section (i.e., a round jet) gives the trivial solution of the
problem. In what follows, we will obtain its nontrivial solutions
corresponding to the azimuthal deformations of the round jet surface. 

For convenience we convert to the dimensionless variables:
\begin{gather*}
x\to Q^2(2\pi\alpha)^{-1}\,x, \qquad
y\to Q^2(2\pi\alpha)^{-1}\,y,
\\
\varphi\to 2Q\,\varphi,
\qquad
p\to 2\pi\alpha^2Q^{-2}\,p.
\end{gather*}
By analogy with Ref.~\cite{cra}, we choose $f=\ln|\nabla\varphi|$ as a new
unknown function, and the pair of conjugate harmonic functions $\varphi$
and $\psi$ as new independent variables (the condition $\psi=\mbox{const}$
defines the electric field lines). 
The so-called complex potential $w=\varphi-i\psi$ is an analytic function of
the complex variable $z=x+iy$. The complex expression $\ln(-dw/dz)$ is
also an analytic function, and, as a consequence, the real functions 
\begin{gather}
f\equiv\mbox{Re}\,\ln(-dw/dz)=\ln|\nabla\varphi|,
\label{st1} \\
\theta\equiv-\mbox{Im}\,\ln(-dw/dz)=
\arctan\left(\varphi_y/\varphi_x\right)
\label{st2} 
\end{gather}
are conjugate harmonic functions of the variables $\varphi$ and $\psi$.
In particular, this implies that the function $f$ satisfies the Laplace
equation:
\begin{equation} f_{\varphi\varphi}+f_{\psi\psi}=0.
\label{stac9}
\end{equation}
The boundary conditions for $f$ can be derived from the
expressions (\ref{stac3}) and (\ref{stac4}). We get
\begin{gather}
f_\varphi=p\,e^{-f}+e^f,
\qquad \varphi=0,
\label{stac10} \\
f\to\varphi, \qquad \varphi\to-\infty,
\label{stac11}
\end{gather}
where we have taken into account that the fluid surface curvature can be
expressed in terms of the functions $f$ and $\varphi$ in the following way:
$C=-(f_\varphi \exp{f})_{\varphi=0}$. Since in the limit $|z|\to\infty$ we
have $w\to-\ln{z}$ for the complex potential and, consequently, a closed
surface corresponds to changing $\psi$ by $2\pi$, we add the condition for
periodicity of $f$ with respect to the variable $\psi$,
\begin{equation}
f(\varphi,\psi)=f(\varphi,\psi+2\pi),
\label{stac12}
\end{equation}
closing the system of equations for the function $f$.

Thus, the problem of finding the steady-state profile of a cylindrical jet 
surface amounts to studying the boundary-value problem 
(\ref{stac9})--(\ref{stac12}) on the half-plane $\varphi\leq 0$.

\section{Exact solutions}

A wide class of particular solutions of the equations
(\ref{stac9})--(\ref{stac12}) was obtained in Ref.~\cite{zu3}.
They are given by the formula
\begin{equation}
f=\ln\!\left(\!\frac{l-1}{2}\!\right)\!
+\ln\!\left(\!
\frac{1+a^2b^2e^{2n\varphi}\!+2abe^{n\varphi}\!\cos(n\psi)}
{a^2+b^2e^{2n\varphi}-2abe^{n\varphi}\cos(n\psi)}\!\right)\!+\varphi,
\label{stac38}
\end{equation}
where we put
$$
a=\sqrt{(n-1)/(n+1)},\quad b=\sqrt{(n-l)/(n+l)}.
$$
In these expressions $n$ is the azimuthal mode number ($n=2,3,4...$) and
$l=\sqrt{1-4p}$ is the parameter characterizing the amplitude of the
surface deformation of the round jet. 

Let us construct the equilibrium surfaces corresponding to the
solution (\ref{stac38}). It follows from the definitions (\ref{st1}) and
(\ref{st2}) that the inverse transformation from the variables $\varphi$
and $\psi$ to $x$ and $y$ is determined by the relation 
\begin{equation}
z=-\!\int\!\exp(-f+i\theta)dw,
\label{stac43}
\end{equation}
where $\theta$ is the angle of inclination of the electric field intensity
vector to the abscissa direction. Taking into account that $f$ and
$\theta$ are conjugate harmonic functions and thus the Cauchy-Riemann
conditions are satisfied ($\theta_\psi=f_\varphi$ and
$\theta_\varphi=-f_\psi$), we obtain from Eq.~(\ref{stac38}):
$$
f-i\theta=\ln\left(\frac{l-1}{2i}\right)+
2\ln\left(\frac{1+abe^{nw}}{a-be^{nw}}\right)+w.
$$
Substituting this expression into (\ref{stac43}), we get:
\begin{equation}
z=\frac{2i}{(l-1)}\int\!
\left(\frac{at^n-b}{t^n+ab}\right)^2\!\!dt
=\frac{2ia^2t\left(t^n+b/a^3\right)}
{(l-1)\left(t^n+ab\right)},
\label{nnn18}\end{equation}
where we have introduced the notation $t=\exp(-w)$. The transformation
corresponding to (\ref{nnn18}) maps the unit circle $|t|=1$ onto the free
surface of a liquid. The algebraic expression (\ref{nnn18}) relates the
results of Ref.~\cite{zu3} and the results of Refs.~\cite{cro,weg}, where
the conformal mapping to the exterior of the init disc was applied for the
problem related to an analysis of the profile of a two-dimensional bubble
in a circulatory ambient flow.

Bearing in mind that $\varphi=0$ at the boundary and, consequently,
$t=\exp(i\psi)$, we obtain from (\ref{nnn18}) that the sought-for
equilibrium surfaces are given by the following parametric expression:
\begin{equation}
z=\frac{2ia^2e^{i\psi}\!\left(e^{in\psi}\!+b/a^3\right)}
{(l-1)\left(e^{in\psi}\!+ab\right)},
\label{sss18}\end{equation}
where $\psi$ plays the role of the parameter. The closed surface
corresponds to the change in $\psi$ in the range $0\leq\psi<2\pi$.

Let us return to the real variables. Having separated the real part
from the imaginary one in (\ref{sss18}), we find: 
\begin{equation}
y=\!\frac{a^{-1}b\cos{(n\psi\!-\!\psi)}\!+\!a^3b\cos{(n\psi\!+\!\psi)}
\!+\!(a^2\!+\!b^2)\cos{\psi}}
{(1+a^2b^2+2ab\cos{(n\psi)})(l-1)/2},
\label{sss20} 
\end{equation}
\begin{equation}
x=\!\frac{a^{-1}b\sin{(n\psi\!-\!\psi)}\!-\!a^3b\sin{(n\psi\!+\!\psi)}
\!-\!(a^2\!+\!b^2)\sin{\psi}}
{(1+a^2b^2+2ab\cos{(n\psi)})(l-1)/2}.
\label{sss21}\end{equation}
The formulas (\ref{sss20}) and (\ref{sss21}) represent a family of exact
two-parametric solutions for the equilibrium shape of a charged jet of a
conducting liquid. To the best of our knowledge, these solutions for the
jet configurations have not been considered so far. 

The particular solutions (\ref{sss20}) and (\ref{sss21}) of the initial
equations relate to the case $p\not=0$. Note that for $p=0$ it is
possible to find the general solution of
Eqs.~(\ref{stac9})--(\ref{stac12}). It has the following form: 
\begin{equation}
f=\ln\left(1-c^2\right)\!
-\ln\left(c^2\!-2ce^{-\varphi}\!\cos{\psi}+e^{-2\varphi}\right)-\varphi,
\label{add2}
\end{equation}
where the constant $c$ satisfies the inequality $0\leq c<1$.
This solution is $2\pi$-periodical with respect to $\psi$,
so that it corresponds to the azimuthal number $n=1$. 
The equilibrium surface is given by the parametric expression 
$$
z=\frac{1}{1-c^2}\,\left(e^{i\psi}-c^2e^{-i\psi}-2c\,i\psi\right).
$$
One can readily see that this surface has the self-intersections for
arbitrary $c$. Thus, the solution (\ref{add2}) with the mode
number $n=1$ is physically meaningless.

\section{Conditions of the jet splitting}

The solutions (\ref{sss20}) and (\ref{sss21}) obtained in
the previous section enable us to find exact critical values of the 
linear charge densities required (i) for the onset of the azimuthal
instability of the round jet and (ii) for the jet splitting.

\begin{figure}
\includegraphics{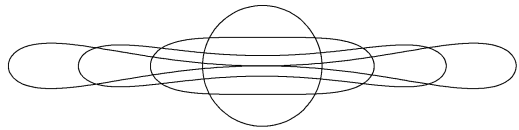}
\caption{\label{fig:fig_2}
Typical (superposed) cross sections of the charged jet of a conducting
liquid for $n=2$ and parameter values $l=1.86, 1.94, 1.88, 2$.
The cross section areas are normalized to a constant.
}
\end{figure}

\begin{figure}
\includegraphics{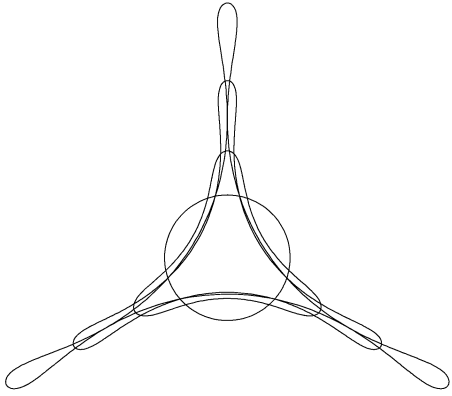}
\caption{\label{fig:fig_3}
The cross sections of the jet for $n=3$ and parameter values 
$l=2.53, 2.59, 2.78, 3$.
}
\end{figure}

\begin{figure}
\includegraphics{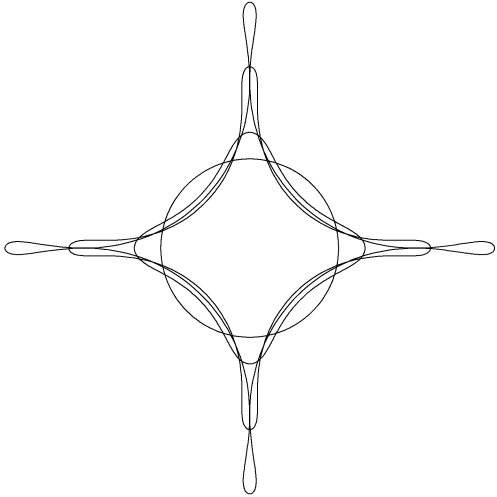}
\caption{\label{fig:fig_4}
The cross sections of the jet for $n=4$ and parameter values 
$l=3.19, 3.35, 3.8, 4$.
}
\end{figure}

In the limit $l\to n$ expressions (\ref{sss20}) and (\ref{sss21}) for the
equilibrium shape of a charged jet of a conducting liquid define circles
of radius $2/(n+1)$, which correspond to the unperturbed state of the
jet, namely, to the round jet. With decreasing the parameter $l$, the jet
surface is deformed. At certain $n$-dependent critical values $l_n$ of the
parameter $l$, the region occupied by the liquid ceases to be simply
connected, and the jet splits (see Fig.~\ref{fig:fig_2}--\ref{fig:fig_4}).
For $1<l<l_n$ the solutions are physically meaningless so that for fixed
$n$ the set of the problem solutions corresponds to the interval $l_n\leq
l\leq n$.

We now determine the critical values of the parameter $l$. For $n=2$ the
condition of the surface self-intersection has the form
$$
x=0, \qquad \psi=\pi/2,
$$
that corresponds to the singularity at the point $x=y=0$. It follows from
this condition that $l_2$ is a root of the quadratic equation
$$ 7\,l_2^2-6\,l_2-13=0.
$$
Only one solution of the equation, $l_2=13/7\approx1.86$, meets the
requirement $1<l_2<2$. The jet splits into two equal parts at this value
of the parameter $l$ (see Fig.~\ref{fig:fig_
2}). 

For $n>2$ the condition of the self-intersection of the surface can be
written as 
$$
x=0, \qquad dx/d\psi=0.
$$
Hence it follows that the critical values of $l$, i.e., the quantities
$l_n$, are determined by the set of equations:
\begin{gather}
a(n+1)\left(a^2+b^2\right)\cos{\psi}\qquad
\nonumber \\
=2b(n-1)
\left[n\sin(n\psi)\sin{\psi}+\cos(n\psi)\cos{\psi}\right],
\label{sss22} 
\end{gather}
\begin{gather}
a(n+1)^2\left(a^2+b^2\right)\sin{\psi}\qquad
\nonumber \\
=2b
\left[2n\sin(n\psi)\cos{\psi}-(n^2+1)\cos(n\psi)\sin{\psi}\right].
\label{sss23}\end{gather}
It is easy to see that, dividing (\ref{sss23}) by (\ref{sss22}), we can
eliminate the parameter $l$ from these equations. As a result we obtain
the equation for $\psi$ at the point of the curve self-intersection:
$$
\tan(n\psi)\left[(n^2-1){\tan}^2\psi-2\right]+2n\,\tan\psi=0.
$$ 
With the help of de Moivre's formula, this trigonometrical equation can be
brought to the algebraic form: 
\begin{equation}
\left[(n^2\!-\!1){h}^2\!-\!2\right]\mbox{Im}(1\!+\!ih)^n
\!+2nh\,\mbox{Re}(1\!+\!ih)^n\!=0,
\label{sss25}\end{equation}
where we put $h=\tan{\psi}$. Solving this equation of the $n$th order, we
can define the value of $h$ and, consequently, of $\psi$ for any $n$. 
At given $\psi$, the sought-for critical values of the parameter $l$ can
be determined from the relation (\ref{sss22}), which can be transformed
into the quadratic equation with respect to the unknown quantity $l_n$. 

For $n=3$ the expression (\ref{sss25}) transforms into the trivial
equation $h^2=1$, from which it follows that the self-intersection takes
place at $\psi=\pi/4$. For this value of the parameter $\psi$, the condition
(\ref{sss22}) reduces to solving the equation
$$ 
17\,l_3^2-18\,l_3-63=0.
$$
Its root satisfying the condition $1<l_3<3$ is
$l_3=(9+24\sqrt{2})/17\approx 2.52.$ If $l=l_3$, the volume occupied by
liquid loses its simple connectivity and one jet splits into four unequal
parts. This situation is illustrated in Fig.~\ref{fig:fig_3}.

For $n=4$ (see Fig.~\ref{fig:fig_4}), we obtain from (\ref{sss25}):
$h^2=5/13$, and (\ref{sss22}) transforms into the equation:
$$
443\,l_4^2-486\,l_4-2957=0.
$$
It follows herefrom that $l_4=(243+370\sqrt{10})/443\approx3.19.$
For the next azimuthal number, $n=5$, we deduce from Eq.~(\ref{sss25}):
$$
3h^4-24h^2+5=0,
$$
and hence $h^2=4\pm\sqrt{43/3}$. Substituting this value of the parameter
$h$ into (\ref{sss22}), we get: $l_5\approx3.85$. In a similar manner, we
can find that $l_6\approx4.55$ and $l_8\approx5.85$. 

Thus, we have defined the values of the parameter $l$ for which the
solutions of the problem of the equilibrium configurations of the jet
surface exist. Let us now consider a jet with given characteristics (the
cross section area $S$ and the surface tension coefficient $\alpha$) and
determine the linear charge densities $Q$, which correspond to the
allowable values of $l$, i.e., to the interval $l_n\leq l\leq n$.  

For the family of the solutions (\ref{sss20}) and (\ref{sss21}), the area
$S$ can be easily found with the help of the Green's formula (compare with
Ref.~\cite{weg}): 
$$
S=-\frac{n}{2}\,\mbox{Im}\!\!\!\int\limits_0^{2\pi/n}\!\!\!
z\,\frac{d\bar{z}}{d\psi}\,d\psi=
\frac{4\pi\left((l+1)^2-4n\right)}{(l^2-1)^2},
$$
where the dependence (\ref{sss18}) of the complex variable $z$ on $\psi$
have been used. Returning to initial dimensional quantities, we get
$$
S=\frac{Q^4\left((l+1)^2-4n\right)}{\pi{\alpha}^2(l^2-1)^2}.
$$
Solving this relation with respect to $Q$, we arrive at the dependence of
the linear charge density on the surface tension $\alpha$, the jet cross
section area $S$, and the steady-state solution parameters $l$ and $n$:
\begin{equation}
Q=\left[\frac{\pi{\alpha}^2S(l^2-1)^2}{(l+1)^2-4n}\right]^{1/4},
\label{sss29}
\end{equation}
It enables us to determine the critical values of the charge density.

In the problem under consideration, we can define two critical charge
density values for every azimuthal wavenumber $n$. The first critical
density value $Q_n$ corresponds to the threshold of linear instability of 
the jet with round cross section. It can be calculated from the formula
(\ref{sss29}), where we must take $l=n$ (recall that for $l=n$ the
expressions (\ref{sss20}) and (\ref{sss21}) define circles):
$$
Q_n=\left[\pi(n+1)^2\alpha^2S\right]^{1/4}.
$$
The critical charge density is seen to grow monotonically with $n$. 
Its minimum value corresponds to $n=2$:
$$
Q_2=\left(9\pi\alpha^2S\right)^{1/4}\approx 2.31\,\alpha^{1/2}S^{1/4}.
$$
If the linear charge density $Q$ of the jet exceeds this value, an
electrohydrodynamic instability of the jet surface will develop. At the
initial stage of the process, this leads to the elliptic deformation of the jet
cross-section. The disturbances corresponding to the modes with
lesser spatial scales, $n>2$, can grow only at larger values of the linear
charge density.  

The second critical density value $\tilde{Q}_n$ corresponds to the
situation when the volume occupied by the jet loses its simple
connectivity and the jet splits into two or more separate jets (see
Fig.~\ref{fig:fig_2}--\ref{fig:fig_4}). We can find it substituting
$l=l_n$ into (\ref{sss29}): 
$$
\tilde{Q}_n=\left[\frac{\pi{\alpha}^2S(l_n^2-1)^2}
{(l_n+1)^2-4n}\right]^{1/4}.
$$
The charge density takes the minimal value for $n=2$, whence it follows that the
jet splitting into two approximately equal parts can be considered as
the most probable scenario of the jet disintegration. Since $l_2=13/7$,
the exact minimal value of the second critical density is given by the
following expression: 
$$
\tilde{Q}_2=\left(1800\,\pi\alpha^2S/49\right)^{1/4}\!\!\!
\approx3.28\,\alpha^{1/2}S^{1/4}.
$$

Note that the critical densities $Q_n$ can be found from the linear
analysis of the stability of the round jet surface (see, for example,
Ref.~\cite{sav,shi}). In so doing it is not necessary to know the exact solutions for
the stationary jet shape. However we must know them for determining the
conditions of the jet splitting or, what is the same, for finding the
second critical charge densities.

\section{Stability analysis}

Now consider the dependence of the jet surface deformation amplitude on
the linear electric charge density. Such an analysis will allow us to draw some
qualitative conclusions concerning the stability of the solutions obtained. 

\begin{figure}
\includegraphics{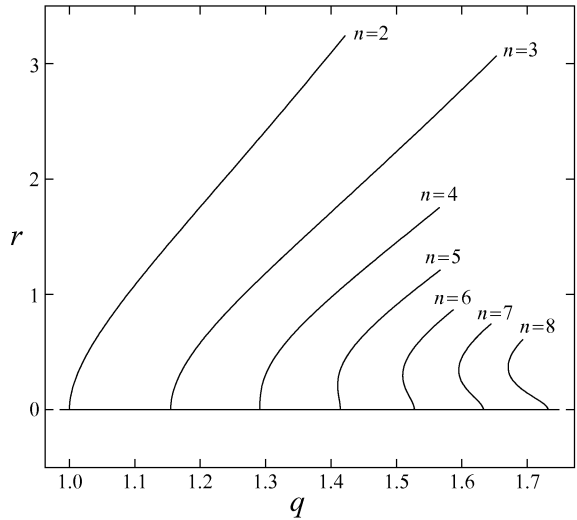}
\caption{\label{fig:fig_5}
The dimensionless amplitude of the jet surface deformation $r$ as a
function of the normalized linear charge density $q$ for different 
azimuthal numbers $n$. The straight line $r=0$ corresponds to the
unperturbed state of the system, i.e., to the round jet.
}
\end{figure}

It is convenient to take $Q_2$ as a unit of linear electric charge density
and the radius of the unperturbed jet $R_0=\sqrt{S/\pi}$ as a unit 
of length. This implies introducing both dimensionless charge density $q$ and
the deformation amplitude of the jet surface $r$: 
$$
q(n,l)\equiv\frac{Q}{Q_2}=
\left[\frac{(l^2-1)^2}{9\left((l+1)^2-4n\right)}\right]^{1/4},
$$
$$
r(n,l)\equiv\frac{R_{\max}\!-R_0}{R_0}=
\frac{2\sqrt{n^2\!-l^2}+(l-\!1)\sqrt{n^2\!-\!1}}
{\sqrt{(n^2-1)\left((l+1)^2-4n\right)}}-1,
$$
where $R_{\max}=y|_{\psi=0}$ is the maximum distance between the jet axis
and its surface. These relations give the dependence of $r$ on $q$ for
different $n$ in the parametric form ($l$ plays the role of the
parameter). The relevant plots are presented in Fig.~\ref{fig:fig_5}. 
The straight line $r=0$ in the figure corresponds to the unperturbed state
of the system, i.e., to the round jet. 

One can notice that the deformation amplitude monotonically increases with
the charge density for $n=2,3,4$. This situation corresponds to the soft
loss of stability of the round jet. Indeed, it is clear that if
$q<Q/Q_n$, then the round jet is stable with respect to small surface
perturbations with azimuthal wavenumbers greater than or equal to $n$.
Bifurcations occur when the condition $q=Q/Q_n$ holds (supercritical
bifurcation for $n=2,3,4$). It is seen from the figure that the side
branches corresponding to our exact solutions fork from the straight line
$r=0$. It should be noted that the authors of Ref.~\cite{weg}, in terms of
the present paper, have plotted the dependence of $q^2$ on $p$ that also
indicates the manner in which the solution branches bifurcate.  

As the trivial solution $r=0$ is unstable for $q>1$,
the free energy of the system can have a minimum on these branches only.
This suggests that the surface modes with small azimuthal wavenumbers
are excited in a soft regime and, at least for a small overcriticality,
our exact solutions can be stable with respect to small perturbations that
do not violate the problem symmetry.

For $n>4$, the surface deformation amplitude $r$ decreases with increase
in $q$ in the vicinity of the branch points (see Fig.~\ref{fig:fig_5}).
Such a dependence of the amplitude $r$ on the control parameter $q$ 
corresponds to the subcritical bifurcation. Then the potential energy of
the system, i.e., the sum of the surface and electric field energies, has
a maximum on the side branches. It immediately follows that
our solutions having azimuthal mode numbers $n>4$ are unstable with respect
to small changes in the amplitude $r$.

Since both the first and second critical charge densities are minimal for the
large-scale azimuthal mode $n=2$, it is the surface mode that will define the jet
behavior. If the linear charge density $Q$ exceeds the value $Q_2$, then
the surface of the cylindrical jet with a round cross section becomes
unstable. As the instability regime is soft, then, for a small
subcriticality, a new stable state corresponding to our stationary
solutions with $n=2$ appears, and the round jet transforms to an elliptic
one. Unfortunately, the analysis of the balance conditions for the forces
acting on the jet surface cannot provide the answer to the question as to
whether our solutions are stable for all permissible values of the
electric charge density, $Q_2<Q<\tilde{Q}_2$ (we discuss the stability
with respect to surface perturbations that do not violate the problem
symmetry). In all probability, they are unstable for a sufficiently large
$Q$. The point is that $Q_3<\tilde{Q}_2$ and $Q_4<\tilde{Q}_2$, and,
consequently, the functional of the jet potential energy can have several
extremums corresponding to different $n$ at given $Q$.  Moreover, as
evident from Fig.~\ref{fig:fig_2}, the jet cannot be stable if the charge
density value is close 
to $\tilde{Q}_2$. An arbitrary small deformation of the surface can lead
to the rupture of the neck. Thus the condition $Q>\tilde{Q}_2$ may be
considered as the sufficient condition for splitting the jet into two.

\section{Concluding remarks}

In the present work we have obtained the two-parameter family of the exact
solutions of the classical problem in electrostatics, namely, the problem
of finding the equilibrium configuration of a charged jet of a conducting
liquid. The approach applied is based on the conformal mapping of the
region outside the jet to the half-plane, that have restricted our
consideration to the case of the plane symmetry of the problem, when all
quantities depend only on two variables $x$ and $y$ (see
Fig.~\ref{fig:fig_1}).  Because of this, all the solutions obtained
correspond to the azimuthal deformations of the jet surface, whereas the
deformations in the direction of the jet axis have not been considered. At
the same time, it is just the longitudinal instabilities (the varicose and
sinuous modes) that determine the behavior of a charged jet in the general
case \cite{sav}. Nevertheless, if the electrohydrodynamic instability of the
liquid cylinder is suppressed in the direction of the $z$ axis, our
solutions can play a dominant role in the jet behavior. For instance, the
growth of the axial disturbances can be stabilized by the magnetic field
directed along the jet axis. It is known \cite{mel} that the tangential
magnetic field retards the development of the surface instabilities which
bend the field lines (in particular, this phenomenon is used to confine
the plasma). If the magnetic field is sufficiently strong, the only
azimuthal instability of the jet surface will develop. As it was discussed
in the previous two sections, such instability leads to the jet splitting.

\medskip

This study was supported by the President of Russian Federation (Project
No. MK-2149.20042), the Foundation for Support of Russian Science, the
``Dynasty'' Foundation and the International Center for Fundamental Physics
in Moscow.

\end{document}